\documentclass[conference]{IEEEtran}
\IEEEoverridecommandlockouts
\usepackage{amsmath,amssymb,amsfonts}
\usepackage{algorithmic}
\usepackage{graphicx}
\usepackage{textcomp}
\usepackage{xcolor}
\usepackage{multirow}
\usepackage{hyperref}
\usepackage[numbers]{natbib}
\usepackage{csquotes}
\usepackage{mathtools}
\usepackage{booktabs}
\usepackage{tikz}
\usetikzlibrary{positioning}

\def\BibTeX{{\rm B\kern-.05em{\sc i\kern-.025em b}\kern-.08em
    T\kern-.1667em\lower.7ex\hbox{E}\kern-.125emX}}

\renewenvironment{thebibliography}[1]{
	\begin{oldthebibliography}{#1}
		 \setlength{\itemsep}{-.3ex}%
	}
	{
	\end{oldthebibliography}
}

\definecolor{color_256693}{rgb}{0.894118,0.968628,0.537255}
\definecolor{color_29791}{rgb}{0,0,0}
\definecolor{color_175461}{rgb}{0.568628,0.768628,0.976471}
\definecolor{color_208162}{rgb}{0.705882,0.654902,0.839216}
\definecolor{color_117641}{rgb}{0.341177,0.505882,0.956863}
\definecolor{color_245272}{rgb}{0.85098,0.85098,0.85098}
\definecolor{color_274846}{rgb}{1,0,0}
\definecolor{color_275101}{rgb}{1,0,1}
\definecolor{color_236339}{rgb}{0.811765,0.933333,0.952941}

\begin{document}

\title{MCDDPM: Multichannel Conditional Denoising Diffusion Model for Unsupervised Anomaly Detection in Brain MRI\\
}

\author{
	Vivek Kumar Trivedi \textsuperscript{1*}, 
    Bheeshm Sharma \textsuperscript{1$\dagger$}, 
    P. Balamurugan \textsuperscript{1$\ddagger$} 

	\vspace{0.5em}\\
	\textsuperscript{1} IEOR, Indian Institute of Technology Bombay, Mumbai, India. 
	\\ 
    \textsuperscript{*}vivekkumartrivedi9@gmail.com,  \textsuperscript{$\dagger$}bheeshmsharma@iitb.ac.in, \textsuperscript{$\ddagger$}balamurugan.palaniappan@iitb.ac.in
}

\maketitle

\begin{abstract}
	Detecting anomalies in brain MRI scans using supervised deep learning methods presents challenges due to anatomical diversity and labor-intensive requirement of pixel-level annotations. Generative models like Denoising Diffusion Probabilistic Model (DDPM) \cite{ho2020denoising} and their variants like Patch-based DDPM (pDDPM) \cite{behrendt2024patched}, Masked DDPM (mDDPM) \cite{iqbal2023unsupervised}, Conditional DDPM (cDDPM) \cite{behrendt2023guided} have recently emerged to be powerful alternatives to perform unsupervised anomaly detection in brain MRI scans. These methods leverage frame-level labels of healthy brains to generate healthy tissues in brain MRI scans. During inference, when an anomalous (or unhealthy) scan image is presented as an input, these models generate a healthy scan image corresponding to the input anomalous scan, and the difference map between the generated healthy scan image and the original anomalous scan image provide the necessary pixel level identification of abnormal tissues. The generated healthy images from the DDPM, pDDPM and mDDPM models however suffer from fidelity issues and contain artifacts that do not have medical significance. While cDDPM achieves slightly better fidelity and artifact suppression, it requires huge memory footprint and is computationally expensive than the other DDPM based models. In this work, we propose an improved version of DDPM called Multichannel Conditional Denoising Diffusion Probabilistic Model (MCDDPM) for unsupervised anomaly detection in brain MRI scans. Our proposed model achieves high fidelity by making use of additional information from the healthy images during the training process, enriching the representation power of DDPM models, with a computational cost and memory requirements  on par with DDPM, pDDPM and mDDPM models. Experimental results on multiple datasets (e.g. BraTS20, BraTS21) demonstrate promising performance of the proposed method. The code is available at \href{https://github.com/vivekkumartri/MCDDPM}{https://github.com/vivekkumartri/MCDDPM}.
\end{abstract}


\section{Introduction}
Medical imaging systems are essential for radiologists, aiding in precise diagnosis and patient care decisions by offering detailed visualizations \cite{lundervold2019overview}. Magnetic Resonance Imaging (MRI) segmentation is crucial to identify anomalies like necrosis and edema \cite{fernando2021deep}, which are often clearly visible on T2-weighted (T2w) images \cite{tschuchnig2022anomaly}. When extensive pixel level labeling is not available, Unsupervised Anomaly Detection (UAD) is useful where several techniques like level sets \cite{taheri2010level}, graph cuts \cite{garcia2009multiple}, AutoEncoders (AEs), Variational AutoEncoders (VAEs) \cite{baur2021autoencoders}, and Generative Adversarial Networks (GANs) \cite{nguyen2021unsupervised} are available. 

Recently, generative models like denoising diffusion probabilistic models (DDPMs) \cite{ho2020denoising} have garnered significant attention for their capability to produce high-quality images by removing noise from corrupted inputs. 
DDPM uses the idea of diffusion \cite{sohl2015deep}, and preserves spatial information in its hidden layer representations, crucial for maintaining the structural integrity, sharpness and high fidelity of the generated images \cite{wyatt2022anoddpm}. Patch-based DDPM (pDDPM) \cite{behrendt2024patched} improves the preservation of anatomical consistency by reconstructing separate patches of the MRI scans, ensuring that particular brain structures maintain their position consistently between input and reconstructed images. Another improved variant called mDDPM  \cite{iqbal2023unsupervised} integrates Masked Image Modeling (MIM) and Masked Frequency Modeling (MFM) techniques within DDPM framework, offering superior generalization abilities. Context-conditioned DDPM (cDDPM) \cite{behrendt2023guided} improves anatomical coherence and aligned intensity characteristics in MRI scans by incorporating latent feature representations of noise-free input images into the denoising process and conditioning the model on contextual information. However, cDDPM requires a separate model for contextual information, leading to increased parameter count and computational complexity. Despite slight improvements, cDDPM too suffers from similar challenges as the other DDPM models in accurate reconstruction of MRI scans.

\textbf{Our Contributions:} In this work, we propose an approach called Multichannel Conditional DDPM (MCDDPM) which addresses the shortcomings of other DDPM models by incorporating multichannel information (latent representations of noisy images derived from a suitable network) and robust conditioning mechanisms. This extra context, provided through multiple channels, improves the accuracy and realism of the final images. The proposed MCDDPM integrates the contextual information directly without the need for separate models and extra trainable parameters, thus simplifying model design. Our experiments on several benchmark datasets show that MCDDPM is efficient and effective in producing high-quality reconstructions, aiding in precise pixel level localization of anomalous regions in brain MRI scans.

\section{Recent Work}
UAD in brain MRI using AutoEncoders (AEs) for reconstruction has been popular; however, blurry reconstructions limit the effectiveness of AEs in UAD \cite{baur2021autoencoders}. Incorporating skip connections with dropout \cite{baur2020bayesian} and employing feature activation maps \cite{silva2022constrained} have shown to improve the quality of both the learned representations and reconstructions of AEs. Furthermore, \cite{behrendt2022unsupervised} proposed online outlier removal for AEs to enhance their UAD performance. Despite this, AEs can fall prey to a \enquote{copy task} phenomenon, where AEs unintentionally reconstruct abnormal anatomies, hindering their UAD efficacy \cite{bercea2023generalizing}. Regularizing AEs with an additional denoising task \cite{kascenas2022denoising} has helped in tackling this copy task issue. 
Variational AutoEncoder (VAE) based approaches with stronger contextual understanding based on 3D MRI data have also emerged popular \cite{bercea2023mask}, along with 
restoration techniques \cite{chen2020unsupervised} and the fusion of VAEs with GANs \cite{pinaya2022unsupervised}. 

Recent advancements in DDPM have shown promise in comprehensive tissue reconstruction in brain MRI \cite{wyatt2022anoddpm}. While DDPM excels in fidelity, it often encounters information loss during the noising process. Patch-based DDPM (pDDPM) \cite{behrendt2024patched} mitigates this drawback by incorporating original image content for reconstruction. However, pDDPM introduces complexity and computational overhead in processing multiple patches of the same image, resulting in artifacts in regions with overlapping patches. Alternatively, conditioning DDPM with input image knowledge has demonstrated efficacy in various tasks \cite{behrendt2023guided}, however at the cost of leveraging a secondary model for latent representation acquisition. Due to these limitations, none of these models alone can completely reconstruct healthy tissue. Our proposed MCDDPM aims to address the  challenges of DDPM variants and avoids reliance on a secondary architecture for conditioning. 

\section{Proposed Methodology}
In this section, we first briefly introduce DDPM architecture, followed by details of our proposed model called MCDDPM.
\subsection{DDPM}
Denoising Diffusion Probabilistic Model (DDPM) serves as a generative model aimed at learning the underlying data distribution. Consider an underlying (unknown) image data distribution $q(x)$ and let \( X_0 \) denote an image sampled from \( q(x) \), where \( X_0 \in \mathbb{R}^{h \times w} \), with \( h \) denoting height and \( w \) representing width. Training DDPM involves a two-step process comprising a forward and backward phase.

In the forward phase, input image \( X_0 \) undergoes iterative noise addition resulting in successive noisy images $X_t$, \( t\in [T] \coloneqq \{1,2,\ldots,T\} \), where \( T \) denotes the total number of time-steps. At \( T \)-th time-step, the image becomes  completely Gaussian noise, hence \( X_T = \epsilon_T \sim \mathcal{N}(0,I) \). Noise addition at time-step \( t \) is  characterized using a Gaussian as follows:
\begin{align}
	q(X_t | X_{t-1}) = \mathcal{N}(X_t; \mu_t = \sqrt{1 - \beta_t} X_{t-1}, \sigma^2_t = \beta_t \mathbf{I}) \nonumber
\end{align}
where \( \beta_t \) represents a predefined variance scheduler. Typically, \( \beta_0 = 10^{-4} \) and \( \beta_T = 0.02 \) \cite{ho2020denoising}, with \( \beta_t = \frac{t}{T} (\beta_T - \beta_0) \). The time-step \( t \) sampled from \( t \sim \text{Uniform}([T]) \) controls the amount of noise. After applying parameterization trick, we obtain:
\begin{align}
	X_t \sim q(X_t | X_0) = \mathcal{N}(X_t; \mu_t = \sqrt{\bar{\alpha}_t}X_0, \sigma^2_t = (1 - \bar{\alpha}_t)\mathbf{I}) \nonumber
\end{align}
where \( \bar{\alpha}_t = \prod_{i=1}^t \alpha_i \) and \( \alpha_t = 1 - \beta_t \).

In the backward phase, the aim is to generate a denoised $\hat{X}_0$ from noisy $X_t$, by sampling from the conditional probability distribution $p_{\theta}(X_0 | X_t) = \mathcal{N} (\mu_{\theta}(X_t, t), \Sigma_{\theta}(X_t, t)).$ Following \cite{ho2020denoising}, $\mu_{\theta}$ is approximated using a U-Net architecture \cite{ronneberger2015u} with learnable parameters denoted by $\theta$, while $\Sigma_{\theta}(X_t, t) = \Sigma(t) = \frac{1 - \alpha_{t-1}}{1 - \alpha_t} \beta_t \mathbf{I}$ is directly computed. Then a suitable variational lower bound (VLB) is formulated, which leads to the reconstruction of noise at time-step $t$ using a reconstruction error \cite{ho2020denoising}: $L_{\text{rec}, \epsilon} = \|\epsilon_t - \epsilon_{\theta}(X_t,t)\|^2$, where $\epsilon_t=X_t - X_0$ and $\epsilon_{\theta}$ is obtained using the U-Net architecture parametrized by $\theta$. Instead of directly predicting the noise, the estimation of $\hat{X}_0 = X_t - \epsilon_t$ is pursued in \cite{behrendt2024patched}, leading to the loss function $L_{\text{rec}} = \|X_0 - \hat{X}_0\|^2$. The major goal of DDPM is to generate images, hence the backward process usually involves refining a random noise vector through incremental backward steps to reduce noise. However, in the context of UAD task at hand, our objective is not to generate images, but to predict brain tissue anatomy based on an input image. As a result, during inference, we directly predict $\hat{X}_0$ from $X_t$ by following the methodology outlined in \cite{behrendt2023guided}. 

\subsection{MCDDPM}\label{mcddpm_method}

Now we illustrate details of our proposed method called Multichannel Conditional DDPM (MCDDPM), which improves DDPM \cite{ho2020denoising} and their variants, by introducing controlled noise into brain MRI imaging data. Typically brain MRI scans are available as 3D volumetric data of shape $\mathbb{R}^{h\times w \times d}$ (e.g. NIfTI volumes \cite{bakas2018identifying}). We propose to use the 2D slices (or frames) of shape $\mathbb{R}^{h\times w}$ sampled from the 3D volumes in our approach. Initially, a slice denoted as $X_0$ is extracted from the volumetric data and using the forward diffusion process, we generate a fully noisy image ($X^z$) of shape $\mathbb{R}^{h \times w}$, where the noise is added to all pixels of $X_0$.  Along with $X^z$, we also generate a patched noisy image using the forward diffusion, where noise is added only to a patch of $X_0$, resulting in $X^p$ of shape $\mathbb{R}^{h \times w}$. Thus before invoking the backward diffusion phase, we have information from three different sources, including the fully noisy image ($X^z$), the original image ($X_0$), and a patched noisy image ($X^p$). To facilitate effective reconstruction of brain anatomy from the noisy data, note that we used patched noisy image $X^p$ in our forward diffusion, inspired from pDDPM \cite{behrendt2024patched}. For this purpose, we designed patches with varying sizes ($h'_k < h$ and $w'_k < w$, for $k \in [1,2,3...K]$) along with their corresponding binary masks $M_k$ of shape $\mathbb{R}^{h \times w}$, to locate specific patch regions within the larger images marked for noise addition. Further $X^z$ is fed into a multichannel bridge network $B_\vartheta$ parametrized by $\vartheta$, which provides a multichannel latent representation $Z$, capturing multiple intermediate representations of $X^z$. 

\begin{figure*}[t]
{%
\centering
\begin{tikzpicture}[scale=1,xshift=30pt,yshift=0pt,x=1pt,y=1pt]
	\path(0pt,0pt);
	\begin{scope}
		\clip
		(-15pt, -8.1191pt) rectangle (485pt, -271.8808pt);
		
		\draw[color_29791,line width=0.561672pt,line join=round,yshift=-76,xshift=-15,fill=color_256693]
		(217.5941pt, -18.727pt) -- (145.4169pt, -18.727pt)
		-- (145.4169pt, -39.179pt) -- (148.1766pt, -39.179pt)
		-- (142.3565pt, -46.4539pt) -- (136.5365pt, -39.179pt)
		-- (139.2961pt, -39.179pt) -- (139.2961pt, -12.6061pt)
		-- (217.5941pt, -12.6061pt) -- cycle;

		\draw[color_29791,line width=0.561672pt,line join=round,fill=color_175461,yshift=-35,xshift=-10]
		(116.0531pt, -210.629pt) -- (116.0531pt, -85.8898pt)
		-- (209.3604pt, -105.8376pt) -- (209.3604pt, -190.6812pt) -- cycle;	
		
		\draw[color_29791,line width=0.561672pt,line join=round,fill=color_175461,yshift=-76,xshift=-25]
		(401.6927pt, -177.7588pt) -- (401.6927pt, -36.1762pt)
		-- (289.6841pt, -64.4927pt) -- (289.6841pt, -149.4423pt) -- cycle;
		
		\draw[color_29791,line width=0.561672pt,line join=round,fill=color_208162,yshift=-76,xshift=-25]
		(247.9185pt, -65.3772pt) -- (273.0035pt, -65.3772pt)
		-- (273.0035pt, -148.5577pt) -- (247.9185pt, -148.5577pt) -- cycle;
		
		\draw[color_29791,line width=0.561672pt,line join=round,fill=color_256693,yshift=-65,xshift=-15]
		(267.061pt, -27.8264pt) -- (267.061pt, -27.8264pt)
		-- (267.061pt, -27.8264pt)
		-- (351.101pt, -27.8264pt)
		-- (351.101pt, -47.1946pt)
		-- (348.3826pt, -47.1946pt)
		-- (353.7169pt, -53.9113pt)
		-- (359.0513pt, -47.1946pt)
		-- (356.333pt, -47.1946pt)
		-- (356.333pt, -22.5944pt)
		-- (267.061pt, -22.5944pt) -- cycle;
		
		\filldraw[color_256693,yshift=-66,xshift=-25][even odd rule]
		(212.399pt, -16.2936pt) .. controls (212.399pt, -14.0302pt) and (214.2338pt, -12.1953pt) .. (216.4973pt, -12.1953pt)
		-- (300.291pt, -12.1953pt) .. controls (301.378pt, -12.1953pt) and (302.4204pt, -12.6271pt) .. (303.189pt, -13.3957pt)
		-- (304.3893pt, -32.6868pt) .. controls (304.3893pt, -34.9502pt) and (302.5544pt, -36.785pt) .. (300.291pt, -36.785pt)
		-- (216.4972pt, -36.785pt) .. controls (214.2338pt, -36.785pt) and (212.3989pt, -34.9502pt) .. (212.3989pt, -32.6868pt) -- cycle;

	\end{scope}
	
	\begin{scope}
		\clip
		(-15pt, -8.1191pt) -- (-15pt, -8.1191pt)
		-- (-15pt, -8.1191pt)
		-- (485pt, -8.1191pt)
		-- (485pt, -271.8808pt)
		-- (-15pt, -271.8808pt)
		-- (-15pt, -8.1191pt) -- cycle
		;
	
		\draw[color_29791,line width=0.561672pt,line join=round,yshift=-76,xshift=-25,fill=color_245272] 
		(406.62911pt, -104.9803pt) -- (421.62911pt, -104.9803pt)
		-- (421.62911pt, -102.2161pt) -- (427.21835pt, -107.7444pt)
		-- (421.62911pt, -113.2727pt) -- (421.62911pt, -110.5085pt)
		-- (406.62911pt, -110.5085pt) -- cycle;
		
		\draw[color_29791,line width=0.561672pt,line join=round,yshift=-56,xshift=-28,fill=color_245272]
		(232.9442pt, -104.9803pt)
		-- (240.6838pt, -104.9803pt)
		-- (240.6838pt, -102.2161pt)
		-- (246.212pt, -107.7444pt)
		-- (240.6838pt, -113.2727pt)
		-- (240.6838pt, -110.5085pt)
		-- (232.9442pt, -110.5085pt) -- cycle
		;
		
		\draw[color_29791,line width=0.561672pt,line join=round,yshift=-96,xshift=-28,fill=color_245272]
		(232.9442pt, -104.9803pt)
		-- (240.6838pt, -104.9803pt)
		-- (240.6838pt, -102.2161pt)
		-- (246.212pt, -107.7444pt)
		-- (240.6838pt, -113.2727pt)
		-- (240.6838pt, -110.5085pt)
		-- (232.9442pt, -110.5085pt) -- cycle
		;
	
    \newcommand{\yshiftorgconcat}{110}

	\draw[color_29791,line width=0.561672pt,line join=round,fill=color_245272,xshift=-8]
	(84.97022pt, -132.654pt - \yshiftorgconcat)
	-- (84.97022pt, -121.4855pt - \yshiftorgconcat)
	-- (82.20609pt, -121.4855pt - \yshiftorgconcat)
	-- (87.73436pt, -115.9572pt - \yshiftorgconcat)
	-- (93.26263pt, -121.4855pt - \yshiftorgconcat)
	-- (90.4985pt, -121.4855pt - \yshiftorgconcat)
	-- (90.4985pt, -132.654pt - \yshiftorgconcat) -- cycle
	;

		\draw[color_29791,line width=0.561672pt,line join=round,,yshift=-76,xshift=-25,fill=color_245272]
		(274.7092pt, -104.9803pt)
		-- (282.4488pt, -104.9803pt)
		-- (282.4488pt, -102.2161pt)
		-- (287.977pt, -107.7444pt)
		-- (282.4488pt, -113.2727pt)
		-- (282.4488pt, -110.5085pt)
		-- (274.7092pt, -110.5085pt) -- cycle
		;
		
		\draw[color_29791,line width=0.561672pt,line join=round,fill=color_245272,xshift=-10,yshift=-39]
		(98.4614pt, -104.9803pt) -- (98.4614pt, -104.9803pt)
		-- (98.4614pt, -104.9803pt)
		-- (106.7221pt, -104.9803pt)
		-- (106.7221pt, -102.2161pt)
		-- (111.7292pt, -107.7444pt)
		-- (106.7221pt, -113.2727pt)
		-- (106.7221pt, -110.5085pt)
		-- (98.4614pt, -110.5085pt) -- cycle
		;
		\draw[color_29791,line width=0.561672pt,line join=round,fill=color_245272,xshift=-10,yshift=-113]
		(98.4614pt, -104.9803pt) -- (98.4614pt, -104.9803pt)
		-- (98.4614pt, -104.9803pt)
		-- (106.7221pt, -104.9803pt)
		-- (106.7221pt, -102.2161pt)
		-- (111.7292pt, -107.7444pt)
		-- (106.7221pt, -113.2727pt)
		-- (106.7221pt, -110.5085pt)
		-- (98.4614pt, -110.5085pt) -- cycle
		;

		\draw[color_29791,line width=0.561672pt,dash pattern=on 1.2619011pt off 0.9464258pt,line join=round,xshift=85,yshift=20]
		(-4.752069pt, -75.8787pt)
		-- (-4.752069pt, -86.2523pt)
		;
		\draw[->, >=stealth, color_29791, line width=1,yshift=15] (80.2, -83) -- ++(0, -1);
		
		\draw[color_29791,line width=0.561672pt,dash pattern=on 1.2619011pt off 0.9464258pt,line join=round,xshift=38,yshift=20]
(-4.752069pt, -75.8787pt)
-- (-4.752069pt, -86.2523pt)
;

\draw[->, >=stealth, color_29791, line width=1,yshift=15] (33.2, -83) -- ++(0, -1);

		\draw[color_29791,line width=1,dash pattern=on 1.5619011pt off 1.3464258pt,line join=round]
		(18, -138)
		-- (10,-138)
		-- (10,-228)
		-- (18, -228)
		;
		\draw[->, >=stealth, color_29791, line width=1] (18, -138) -- ++(1, 0);
		\draw[->, >=stealth, color_29791, line width=1] (18, -228) -- ++(1, 0);
		\node[rotate=90] at (4, -180) {Bridge network};
		\draw[color_274846,line width=1.123344pt,dash pattern=on 4.493375pt off 3.370031pt,line join=round]
		(426.7115pt, -269.431pt) -- (426.7115pt, -269.431pt)
		-- (426.7115pt, -269.431pt)
		-- (110.75508pt, -269.431pt);

		\draw[color_274846,line width=1.123344pt]
		(114.61054pt, -267.148pt)
		-- (109.76161pt, -269.1478pt)
		-- (114.61054pt, -271.1433pt) -- cycle
		;
		\draw[color_274846,line width=1.123344pt,dash pattern=on 4.493375pt off 3.370031pt,line join=round,yshift=-76,xshift=-25]
		(456.7958pt, -154.6843pt)
		-- (456.7292pt, -194.4005pt)
		;
		
		\draw[color_274846,line width=1.123344pt,yshift=-76,xshift=-25]
		(458.6513pt, -152.6854pt) -- (458.6513pt, -152.6854pt)
		-- (458.6513pt, -152.6854pt)
		-- (456.7989pt, -147.5865pt)
		-- (454.9404pt, -152.6832pt) -- cycle
		;
		\node[draw,fill=color_236339, rounded corners=5pt, inner sep=6pt, align=center,font=\small,xshift=30] at (225,-35) (box) {
			    \begin{tabular}{rp{150pt}}
				 & Concatenate \\
				$L_B$ & Loss Corresponding to Bridge network \\
				$L_u$ & Loss Corresponding to U-Net \\
				CA & cross-attention
			\end{tabular}
		};
		
    	\draw[fill=color_117641,xshift=10,yshift=-30] (9.562445,-163.7732) rectangle (34.562445,-193.7732);
		\draw[fill=color_117641,xshift=10,yshift=-30] (12.562445,-166.7732) rectangle (37.562445,-196.7732);
		
\draw[color_29791,line width=0.561672pt,line join=round,fill=color_245272,xshift=-55,yshift=70]
(84.97022pt, -194.2394pt)
-- (84.97022pt, -204.1883pt)
-- (82.20609pt, -204.1883pt)
-- (87.73436pt, -209.7166pt)
-- (93.26263pt, -204.1883pt)
-- (90.4985pt, -204.1883pt)
-- (90.4985pt, -194.2394pt) -- cycle
;

\draw[color_29791,line width=0.561672pt,line join=round,fill=color_245272,xshift=-55,yshift=17]
(84.97022pt, -194.2394pt)
-- (84.97022pt, -204.1883pt)
-- (82.20609pt, -204.1883pt)
-- (87.73436pt, -209.7166pt)
-- (93.26263pt, -204.1883pt)
-- (90.4985pt, -204.1883pt)
-- (90.4985pt, -194.2394pt) -- cycle
;

\draw[color_29791,line width=0.561672pt,line join=round,fill=color_245272,xshift=-55,yshift=-34]
(84.97022pt, -194.2394pt)
-- (84.97022pt, -204.1883pt)
-- (82.20609pt, -204.1883pt)
-- (87.73436pt, -209.7166pt)
-- (93.26263pt, -204.1883pt)
-- (90.4985pt, -204.1883pt)
-- (90.4985pt, -194.2394pt) -- cycle
;

\draw[fill=color_117641,xshift=10,yshift=-30] (9.562445,-110.7732) rectangle (34.562445,-140.7732);
\draw[fill=color_117641,xshift=10,yshift=-30] (12.562445,-113.7732) rectangle (37.562445,-143.7732);

 \newcommand{\yshiftpatchconcatenate}{69}
\newcommand{\xshiftpatchconcatenate}{-8}

\draw[color_29791,line width=0.561672pt,line join=round,fill=color_245272]
(84.97022pt+\xshiftpatchconcatenate, -192.2394pt+\yshiftpatchconcatenate)
-- (84.97022pt+\xshiftpatchconcatenate, -204.1883pt+\yshiftpatchconcatenate)
-- (82.20609pt+\xshiftpatchconcatenate, -204.1883pt+\yshiftpatchconcatenate)
-- (87.73436pt+\xshiftpatchconcatenate, -209.7166pt+\yshiftpatchconcatenate)
-- (93.26263pt+\xshiftpatchconcatenate, -204.1883pt+\yshiftpatchconcatenate)
-- (90.4985pt+\xshiftpatchconcatenate, -204.1883pt+\yshiftpatchconcatenate)
-- (90.4985pt+\xshiftpatchconcatenate, -192.2394pt+\yshiftpatchconcatenate) -- cycle
;

\draw[color_29791,line width=0.561672pt,line join=round,xshift=-133.8,yshift=-9,fill=color_245272]
(207.7655pt, -149.1757pt) -- (213.2938pt, -143.6474pt) -- (218.8221pt, -149.1757pt) -- (216.058pt, -149.1757pt) 
-- (216.058pt, -200.6303pt) -- (218.8221pt, -200.6303pt) -- (213.2938pt, -206.1586pt) -- (207.7655pt, -200.6303pt)
-- (210.5297pt, -200.6303pt)
 --(210.058pt, -176.1757pt)
 --(175.058pt, -176.1757pt)
 --(175.058pt, -170.1757pt)
 --(210.058pt, -170.1757pt) 
 -- (210.5297pt, -149.1757pt) -- cycle;
		
	\end{scope}
	
	\begin{scope}
		\clip
		(-15pt, -8.1191pt) -- (-15pt, -8.1191pt)
		-- (-15pt, -8.1191pt)
		-- (485pt, -8.1191pt) 
		-- (485pt, -271.8808pt)
		-- (-15pt, -271.8808pt)
		-- (-15pt, -8.1191pt) -- cycle
		;		
	\end{scope}
	\begin{scope}
		\clip
		(-15pt, -8.1191pt) -- (-15pt, -8.1191pt)
		-- (-15pt, -8.1191pt)
		-- (485pt, -8.1191pt)
		-- (485pt, -271.8808pt)
		-- (-15pt, -271.8808pt)
		-- (-15pt, -8.1191pt) -- cycle
		;
		\draw[color_29791,line width=1.685016pt,dash pattern=on 6.740062pt off 5.055047pt,line join=round,yshift=10]
		(-10.91351pt, -20.3265pt) -- (-10.91351pt, -20.3265pt)
		-- (-10.91351pt, -20.3265pt)
		-- (140.7344pt, -20.3265pt)
		-- (140.7344pt, -64.449pt)
		-- (-10.91351pt, -64.449pt) -- cycle
		;
		
		\node[font=\fontsize{8}{10}\selectfont,text width=145pt,yshift=15] at (65.410445pt,-36.88775pt) {Image after the forward pass of Diffusion Model with different noise levels:};
		\node[font=\fontsize{8}{10}\selectfont,text width=145pt,yshift=15] at (65.410445pt,-56.38775pt) 
		{\begin{itemize}
				\setlength\itemsep{0pt}
				\item Patch based noisy image
				\item Fully noisy image
		\end{itemize}};
		
		\draw[color_29791,line width=1.123344pt,line join=round,xshift=-8,yshift=-39]
		(83.69135pt, -107.7444pt) -- (91.77588pt, -107.7444pt)
		(87.73361pt, -103.7021pt) -- (87.73361pt, -111.7867pt)
		(83.69135pt, -107.7444pt) .. controls (83.69135pt, -105.5119pt) and (85.50114pt, -103.7021pt) .. (87.73363pt, -103.7021pt)
		.. controls (88.80571pt, -103.7021pt) and (89.83386pt, -104.128pt) .. (90.59195pt, -104.8861pt)
		.. controls (91.35002pt, -105.6442pt) and (91.77589pt, -106.6723pt) .. (91.77589pt, -107.7444pt)
		.. controls (91.77589pt, -109.9769pt) and (89.96609pt, -111.7867pt) .. (87.73363pt, -111.7867pt)
		.. controls (85.50114pt, -111.7867pt) and (83.69135pt, -109.9769pt) .. (83.69135pt, -107.7444pt) -- cycle
		;
		
		\draw[color_29791,line width=1.123344pt,line join=round,xshift=-8,yshift=0]
		(83.69135pt, -220.3824pt) -- (91.77588pt, -220.3824pt)
		(87.73361pt, -216.3402pt) -- (87.73361pt, -224.4247pt)
		(83.69135pt, -220.3825pt) .. controls (83.69135pt, -218.15pt) and (85.50114pt, -216.3402pt) .. (87.73363pt, -216.3402pt)
		.. controls (88.80571pt, -216.3402pt) and (89.83386pt, -216.7661pt) .. (90.59195pt, -217.5242pt)
		.. controls (91.35002pt, -218.2822pt) and (91.77589pt, -219.3104pt) .. (91.77589pt, -220.3825pt)
		.. controls (91.77589pt, -222.6149pt) and (89.96609pt, -224.4247pt) .. (87.73363pt, -224.4247pt)
		.. controls (85.50114pt, -224.4247pt) and (83.69135pt, -222.615pt) .. (83.69135pt, -220.3825pt) -- cycle
		;

	\end{scope}
	
	\begin{scope}
		\clip
		(-15pt, -8.1191pt) -- (-15pt, -8.1191pt)
		-- (-15pt, -8.1191pt)
		-- (485pt, -8.1191pt)
		-- (485pt, -271.8808pt)
		-- (-15pt, -271.8808pt)
		-- (-15pt, -8.1191pt) -- cycle
		;

		\draw[color_29791,line width=1.123344pt,dash pattern=on 4.493375pt off 3.370031pt,line join=round,yshift=-76,xshift=-25]
		(345.6884pt, -46.4846pt) -- (345.6884pt, -46.4846pt)
		-- (345.6884pt, -46.4846pt)
		-- (345.6884pt, -41.585pt)
		-- (172.3397pt, -41.585pt)
		-- (172.3397pt, -53.4947pt);
		\draw[color_29791,line width=1.123344pt,yshift=-76,xshift=-25]
		(345.6884pt, -46.4846pt) -- (345.6884pt, -46.4846pt)
		-- (345.6884pt, -46.4846pt)
		-- (344.4251pt, -45.2214pt)
		-- (345.6884pt, -48.6922pt)
		-- (346.9517pt, -45.2214pt) -- cycle
		;
		
		\draw[color_29791,line width=1.123344pt,line join=round,xshift=-149,yshift=177]
		(315.4751pt, -195.4188pt) .. controls (315.4751pt, -193.1863pt) and (317.2849pt, -191.3765pt) .. (319.5174pt, -191.3765pt)
		-- (319.5174pt, -191.3765pt) .. controls (320.5894pt, -191.3765pt) and (321.6176pt, -191.8024pt) .. (322.3757pt, -192.5605pt)
		-- (322.3757pt, -192.5605pt) .. controls (323.1338pt, -193.3185pt) and (323.5597pt, -194.3467pt) .. (323.5597pt, -195.4188pt)
		-- (323.5597pt, -195.4188pt) .. controls (323.5597pt, -197.6512pt) and (321.7499pt, -199.4611pt) .. (319.5174pt, -199.4611pt)
		-- (319.5174pt, -199.4611pt) .. controls (317.2849pt, -199.4611pt) and (315.4751pt, -197.6513pt) .. (315.4751pt, -195.4188pt) -- cycle
		(315.4751pt, -195.4188pt) -- (323.5597pt, -195.4188pt)
		(319.5174pt, -191.3765pt) -- (319.5174pt, -199.4611pt)
		;

	\end{scope}
	
	\node at (426.2011,-182.3985) {\includegraphics[width=39.90081pt,height=38.01681pt]{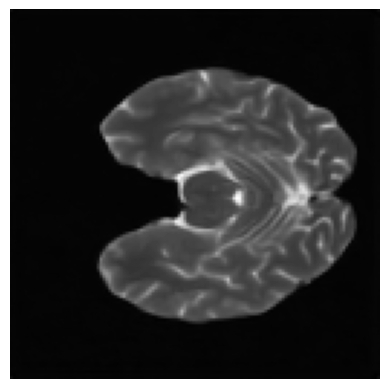}};
	\node at (33.562445,-263.7732) {\includegraphics[width=39.90081pt,height=38.01681pt]{healthy_image.png}};
	\node at (33.562445,-102.7732) {\includegraphics[width=39.90081pt,height=38.01681pt]{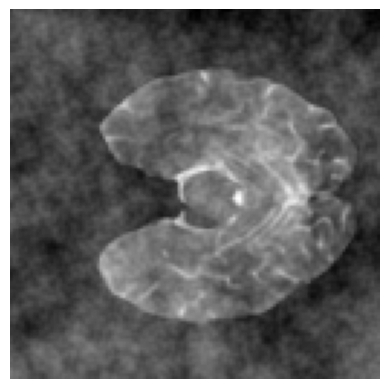}};
	\node at (82.562445,-102.7732) {\includegraphics[width=39.90081pt,height=38.01681pt]{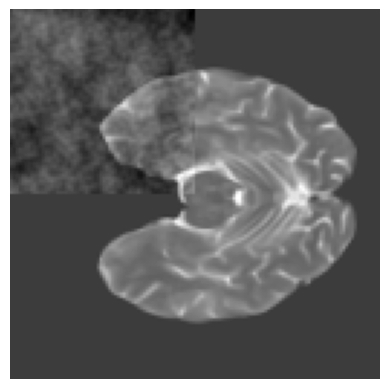}};
	\node at (82.562445,-263.7732) {\includegraphics[width=39.90081pt,height=38.01681pt]{healthy_image.png}};
    \node[text width=75, align=center] at (45, -295) {
	\fontsize{7}{8}\selectfont\color{black}
	\parbox{75pt}{\linespread{0.8}\selectfont Generated Healthy Image by bridge network}
};

    \node[text width=50, align=center] at (88, -75) {
	\fontsize{7}{8}\selectfont\color{black}
	\parbox{50pt}{\linespread{0.8}\selectfont Patch based noisy image}
};

    \node[text width=50, align=center] at (40, -75) {
	\fontsize{7}{8}\selectfont\color{black}
	\parbox{50pt}{\linespread{0.8}\selectfont Fully noisy image}
};
		\draw[color_275101,line width=1.123344pt,dash pattern=on 4.493375pt off 3.370031pt,line join=round,yshift=-305,xshift=40]
(0, 0) -- (0, -10)
-- (50, -10)
-- (50, 5);

		\draw[color_275101,line width=1.123344pt,yshift=-146,xshift=-367]
		(458.6513pt, -152.6854pt) -- (458.6513pt, -152.6854pt)
		-- (458.6513pt, -152.6854pt)
		-- (456.7989pt, -147.5865pt)
		-- (454.9404pt, -152.6832pt) -- cycle
		;
		
		\draw[color_275101,line width=1.123344pt,yshift=-152,xshift=-417]
		(458.6513pt, -152.6854pt) -- (458.6513pt, -152.6854pt)
		-- (458.6513pt, -152.6854pt)
		-- (456.7989pt, -147.5865pt)
		-- (454.9404pt, -152.6832pt) -- cycle
		;

	\node at (152.6189,-184.1155) {\fontsize{18}{1}\selectfont\color{black} Encoder($U_\eta)$};
	\node at (236.7711,-184.1155) {\fontsize{15}{1}\selectfont\color{black} CA};
	\node at (321.7474,-184.1155) {\fontsize{18}{1}\selectfont\color{black} Decoder($U_\phi)$};
	\node at (235.5535,-112.0553) {\fontsize{9}{1}\selectfont\color{black} Skip Connection};
	\node at (235.7466,-92.1029) {\fontsize{9}{1}\selectfont\color{black} Time Embedding};
	\node at (85.905994,-290.0644) {\fontsize{7}{1}\selectfont\color{black} Original Image};
	\node at (35.905994,-212.0769) {\fontsize{18}{1}\selectfont\color{black} $B_\xi$};
	\node at (35.905994,-161.0319) {\fontsize{18}{1}\selectfont\color{black} $B_\vartheta$};
	\node at (234.4057,-229.2803) {\fontsize{9}{1}\selectfont\color{black} Attention Block};
	\node at (425.0072,-209.8882) {\fontsize{7}{1}\selectfont\color{black} Generated Healthy};
	\node at (425.0072,-217.4586) {\fontsize{7}{1}\selectfont\color{black} Image by U-Net};
	\node at (70.83926,-306.1387) {\fontsize{18}{1}\selectfont\color{color_275101} $L_B$};
	\node at (420.293,-249.4522) {\fontsize{18}{1}\selectfont\color{color_274846} $L_u$};
	
\end{tikzpicture}
	\caption{The architecture of the backward pass of MCDDPM.}
	\label{fig:mddpm_architecture}}
	\vspace{-0.2in}
\end{figure*}

Our multichannel diffusion process leverages a U-Net architecture \cite{ronneberger2015u} for image reconstruction. We propose a conditioning approach to enhance the performance of DDPM by integrating a context vector \( C \) (construction of $C$ will be described later) into the backward process. This integration is realized through a modification of the denoising U-Net architecture within the DDPM framework. Specifically, we substitute the bottleneck self-attention layer in the U-Net with a cross-attention (CA) layer (see Attention Block in Fig. \ref{fig:mddpm_architecture}), enabling the model to attend to both information-rich intermediate representations of input data and other contextual information. To facilitate this modification, we partition the U-Net into two segments: the downsampling steps leading up to the bottleneck stage (\( U_{\eta}(.) \)) and the steps following the bottleneck stage (\( U_{\phi}(.) \)). By doing so, we can delineate the processing steps before and after integrating the context vector.

Now we are ready for the backward diffusion process. First, the contextual information is obtained in the following manner. The original clean image $X_0$ is concatenated with the multichannel latent representation $Z$ and fed into the encoder $U_\eta$, along with the time embedding corresponding to zero-th time-step (associated with clean image $X_0$), leading to encoding of the form $U_\eta(X_0\oplus Z, t_0)$. This encoding serves as the contextual information from the clean image and will be subsequently employed for the cross-attention. Next, the patched noisy image $X^p_t$ is concatenated with $Z$ and fed into the encoder $U_\eta$ to yield encoding of the form $U_\eta(X^p_t \oplus Z, t)$, where appropriate time embedding is used in the encoder, corresponding to the time step $t$ used in noise addition during the forward diffusion process to create $X^p_t$. Note that the multichannel information from two different sources $X_0\oplus Z$ and $X^p_t \oplus Z$, corresponding respectively to the clean image $X_0$ and the patched noisy image $X^p_t$ provide the encoder with sufficiently rich information, which can be useful for learning effective structural information of tissue anatomy of brain MRI images. The context vector $C$ is then taken to be the contextual encoding $U_{\eta}(X_0\oplus Z, t_0)$. Subsequently, a cross-attention is performed on the context vector $C$ and  $U_\eta(X^p_t \oplus Z, t)$ to extract correlations between the different encodings. This cross-attended information is passed to the decoder $U_\phi$ to decode the reconstructed clean image $\hat{X}_0 \sim p(X_T) \prod_{t=1}^T p_{\theta}(X_{t-1}|X_t,C)$. Thus the entire backward process can be represented as:   $\hat{X}_0 = U_{\phi}(\text{CA}(U_{\eta}(X_t^p\oplus Z, t),C))$ where \( C \sim U_{\eta}(X_0 \oplus Z, t_0) \).

To enhance the bridge network $B_\vartheta$'s representational power, we feed the latent representation \( Z \) obtained from the completely noisy image $X^z$ into another residual network $B_\xi$, tasked with constructing $\hat{X}^{z} \in \mathbb{R}^{h \times w}$ that retains the original characteristics and dimensions of $X_0$.  Thus the overall loss for the proposed MCDDPM architecture consists of two components: a loss function for the reconstruction obtained at the decoder of U-Net and another for the reconstruction obtained at $B_\xi$ (denoted by $L_u$ and $L_B$ in Fig. \ref{fig:mddpm_architecture}):
\begin{align}
	\mathcal{L} = \| U_{\phi}(\text{CA}(U_{\eta}(X^p_t \oplus Z, t), C)) - X_0 \|_p +\lambda \| \hat{X}^{z} - X_0 \|_p \nonumber
\end{align}
where $\| \cdot \|_p$ denotes $\ell_p$ norm. This dual-term loss function ensures that the model captures both the contextual information through cross-attention and the essential features through the bridge network, enhancing the fidelity of the reconstructed image. The hyperparameter \(\lambda\) allows for balancing the importance of each term, providing flexibility in optimizing the reconstruction process tailored to specific applications or datasets.

\begin{table}[!t]
	\centering
	\caption{Information about Datasets.}
	\resizebox{\columnwidth}{!}{
		\begin{tabular}{lllll}
			\toprule
			\multirow{2}{*}{\textbf{Characteristics}} &\multicolumn{4}{c}{\textbf{Datasets}} \\
			\cmidrule{2-5} 
			&\multirow{1}{*}{\textbf{IXI}} & \multirow{1}{*}{\textbf{BraTS20}} & \multirow{1}{*}{\textbf{BraTS21}} & \multirow{1}{*}{\textbf{MSLUB}}  \\
			\midrule 
			\# Train & 358 & 0  & 0     & 0  \\
			\# Val & 44   & 0 & 100     & 10  \\
			\# Test & 158 & 369    & 1151   & 20 \\
			\midrule
			\textbf{Parameters:} &  &  &  &  \\
			Scanner & Diverse & Diverse & Same & Diverse \\
			Resolution (mm$^3$) & $0.94^{2}$ $\times$ 1.25 & $1^3$ & $0.57^2$ $\times$ 3 & $1^3$ \\
			In-plane size & 256 $\times$ 256 & 240 $\times$ 240 & 240 $\times$ 240 & 240 $\times$ 240 \\
			Z-axis size & 28--136 & 155   & 155 & 155 \\
			Skull-stripped & No    & Yes   & Yes & Yes \\
			\bottomrule
		\end{tabular}%
	}
	\label{tab:dataset_info}%
	\vspace{-0.2in}
\end{table}%

\section{Experimental Setup} \label{experimental_setup}
In this section, we describe the datasets used and the data pre-processing and post-processing procedures used in our experiments. 
\subsection{Datasets}
In this study, we leveraged four publicly available datasets to thoroughly evaluate our proposed methodology. Firstly, for training our proposed model, we utilized the IXI Dataset \cite{IXI}, which consists of 560 pairs of T1 and T2-weighted 3D volumes of brain MRI scans sourced from various institutions. Notably, this dataset exclusively comprises images of only healthy brains, making it particularly suitable for our training purposes. For evaluating the effectiveness of our approach, we integrated three additional datasets. The first one, namely the BraTS20 dataset \cite{bakas2018identifying}, comprises brain MRI scans from patients diagnosed with brain tumors, namely gliomas. This dataset contains 369 3D brain MRI volumes and varies in clinical protocols and scanners from $n=19$ different institutions. Secondly, we used the BraTS21 dataset \cite{baid2021rsna}, which is a widely used resource in the field, encompassing 2040 3D brain MRI scans obtained from patients diagnosed with gliomas. Importantly, both BraTS20 and BraTS21 datasets include detailed annotations in the form of pixel-level binary segmentation masks from expert neuro-radiologists, delineating tumor sub-regions with four different weightings (T1, T1-CE, T2, FLAIR). Thirdly, we utilized the multiple sclerosis dataset from the University Hospital of Ljubljana (MSLUB) \cite{lesjak2018novel}, which includes 3D brain MRI scans from 30 patients diagnosed with multiple sclerosis. Each scan in this dataset is accompanied by T1, T2, and FLAIR-weighted images, with ground truth annotations derived from multi-rater consensus. From all these datasets, we used scans corresponding only to T2 modality. Dataset information is provided in Table \ref{tab:dataset_info} where train, validation and test data splits were done randomly and the 3D MRI volumes in these splits correspond to distinct patients. The validation split was used for hyperparameter tuning. For BraTS20 dataset, we used the same hyperparameters as that for BraTS21 dataset. Additionally, Table \ref{tab:dataset_info} presents a comprehensive overview of various image parameters, such as resolution, image size, Z-axis size, skull stripping and the diversity in scanners used across the datasets.

\subsection{Pre-processing}

In the pre-processing stage, specific procedures were implemented to ensure the consistency and quality of the imaging data obtained from different sources. The IXI dataset \cite{IXI} exhibits variations across MRI centers, hence measures were taken to standardize resolution and orientation. Initially, the resolution was adjusted to $1.0 \times 1.0 \times 1.0$ and orientation to right anterior inferior (RAI) using B-Spline interpolation techniques. Due to the absence of skull stripping in the dataset, the HD-BET \cite{isensee2019automated} architecture was used to perform this task efficiently. Following skull stripping, each volume underwent affine transformation to align it with the T2 modality volume of SRI24-Atlas \cite{rohlfing2010sri24}, enhancing data compatibility. To further enhance data quality, non-relevant black regions within the volumes were eliminated, and N4 Bias field correction \cite{tustison2010n4itk} was systematically applied to mitigate noise effects. For the BraTS20 \cite{bakas2018identifying}, BraTS21 \cite{baid2021rsna} and MSLUB \cite{lesjak2018novel} datasets, wherein skull stripping and registration had already been conducted, similar pre-processing steps were employed, including the removal of non-relevant black regions and the application of N4 Bias field correction for noise reduction. Moreover, for computational efficiency, the volume resolution was reduced by half, resulting in dimensions of [$96 \times 96 \times 80$] voxels. Additionally, to streamline the data, 15 slices from both the top and bottom, parallel to the transverse plane, were removed.

\subsection{Implementation Details}

In this study, all models are implemented using Pytorch (v2.1.2), with data handling and augmentation facilitated by torchio \cite{perez2021torchio}. Intensity ranges are resampled between the 1st and 99th percentiles. During training, time steps are uniformly sampled from the interval $[1, T]$, where $T$ is set to $1000$, while at test time, a fixed time step of $t_{\text{test}} = \frac{T}{2} = 500$ is chosen. The model incorporates a linear schedule for $\beta_t$, ranging from $10^{-4}$ to $2 \times 10^{-2}$. Training proceeds for a maximum of $1600$ epochs, where an epoch is defined based on processing one slice per volume randomly. The best-performing model checkpoint was selected based on the performance on the validation set containing only healthy images for BraTS21 and MSLUB datasets. However for BraTS20 dataset, we used the same checkpoint model obtained for BraTS21 dataset. Volumes are processed slice-wise, with slices uniformly sampled with replacement during training. Utilizing an NVIDIA RTX A5000 (24GB) GPU, training is performed using Adam  optimizer \cite{kingma2014adam}, a learning rate of $10^{-5}$, and a batch size of 8. For the loss function, we utilize the formulation described in \ref{mcddpm_method}. For the patched noisy image $X^p$, we employed $k=4$, resulting in the patched shape $h'_k \times w'_k$ as $48 \times 48$. For the multichannel bridge network (both $B_\vartheta$ and $B_\xi$), we used two residual blocks, with each residual block being the same as described in \cite{he2016identity}. For pDDPM \cite{behrendt2024patched}, cDDPM \cite{behrendt2023guided} and mDDPM \cite{iqbal2023unsupervised} models, the same setup as mentioned in their official GitHub repositories is adopted.

\subsection{Post-processing}

During the inference phase of anomaly detection in medical imaging, the input volume $V$, encompassing both healthy and unhealthy regions, is subjected to a reconstruction process, resulting in the reconstructed volume $\hat{V}$. Following this step, anomaly maps $E_V$ are computed by measuring the residual between the input volume $V$ and its reconstructed counterpart $\hat{V}$ ($\|E_V\|_p = \|V - \hat{V}\|_p$), where higher values of $\|E_V\|_p$ indicate greater reconstruction errors. To mitigate false positives, a series of post-processing techniques are deployed on $E_V$, based on \cite{behrendt2024patched}. Initially, a $5 \times 5 \times 5$ median filter is employed to eliminate minor residual regions, followed by a three-iteration erosion procedure using a brain mask derived from binarizing the input volume $V$. Subsequently, a binary map is obtained by thresholding the residual. In previous methods such as pDDPM \cite{behrendt2024patched}, cDDPM \cite{behrendt2023guided} and mDDPM \cite{iqbal2023unsupervised}, the optimal threshold for segmentation was ascertained using a greedy search strategy on the validation dataset. However, it is improper to utilize segmentation masks from the validation data in an unsupervised setup. Hence, a fixed threshold from the set $\{0.1, 0.2, 0.3, 0.4, 0.5\}$ was chosen, with results reported for the best threshold for all methods, consistently found to be $0.2$.

\section{Results}
In Table \ref{tab:result}, a comprehensive evaluation of segmentation methods across BraTS20, BraTS21 and MSLUB datasets is presented, emphasizing two primary metrics: the Dice coefficient and area under the precision-recall curve (AUPRC). Evaluation includes recent methodologies like DDPM \cite{ho2020denoising}, pDDPM \cite{behrendt2024patched}, cDDPM \cite{behrendt2023guided} and mDDPM \cite{iqbal2023unsupervised}, initially trained on the IXI dataset for generation-based segmentation. These methods were trained on healthy brain images and tested on datasets containing both healthy and unhealthy brain images. The evaluation setup for existing methods was consistent with their official repositories, and for MCDDPM, two loss functions corresponding to the $p$-norm values of $p=1$ and $p=2$, and $\lambda=0.5$ were employed, providing a comparative analysis. For IXI dataset, the reconstruction error obtained by all methods are reported. 

\begin{table}[!h]
	\centering
	\caption{Comparative results}
	\resizebox{\columnwidth}{!}{%
		\begin{tabular}{lcllllll}
			\toprule
			&\multicolumn{7}{c}{\textbf{Datasets}} \\      
			\cmidrule{2-8}
			& \multicolumn{1}{c}{IXI} & \multicolumn{2}{c}{BraTS20} & \multicolumn{2}{c}{BraTS21} & \multicolumn{2}{c}{MSLUB}  \\
			\cmidrule{2-8}
			\textbf{Methods}  & Reconstruction & Dice & AUPRC & Dice & AUPRC & Dice & AUPRC \\
			& Error & & & & & & \\
			\midrule
			DDPM \cite{ho2020denoising} & 0.01346  & 22.16 & 26.77 & 31.36 & 40.11  & 4.39 & 7.44\\
			pDDPM \cite{behrendt2024patched} & 0.01571  & 25.94 & 30.04 & 39.49 & 47.41 & 4.11 & 7.42\\
			cDDPM \cite{behrendt2023guided} & 0.01486  & 28.54 & 33.48 & 41.46 & 50.39 & 5.61 & 5.38\\
			mDDPM \cite{iqbal2023unsupervised} & 0.01763 & 29.22 & 36.52 & 43.6  & 50.14 & 5.32 & 5.34\\
			\midrule
			MCDDPM($p=1$) & 0.01136  & 36.81 & 41.1 & 49.25 & 55.98  & 5.38 & 6.38 \\
			MCDDPM($p=2$) & 0.01132  & \textbf{37.18} & \textbf{41.72} &\textbf{50.56} &\textbf{57.83}  & 5.49 & 6.67 \\
			\bottomrule
		\end{tabular}%
	}
	\label{tab:result}%
\end{table}%
\begin{figure}[t]
	\centering
	\includegraphics[scale=0.5, trim=0 0 0 0, clip]{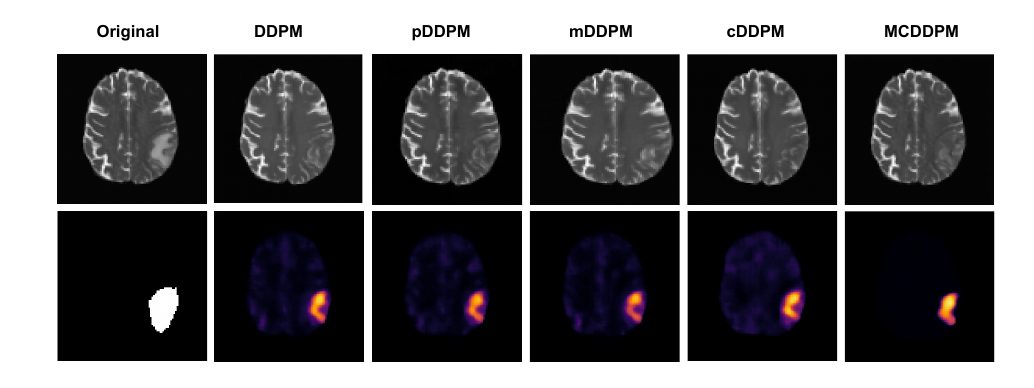}
	\caption{Comparative Qualitative Results. The first column showcases the original image (top row) alongside its corresponding segmentation mask (bottom row). Subsequent columns illustrate the performance of various methods for image reconstruction and anomaly detection. Each row within these columns presents the reconstructed image (top) and the anomalies detected by the respective model (bottom).}
	\label{fig:qualitativeresults}
	\vspace{-16pt}
\end{figure}

The summarized results, as shown in Table \ref{tab:result}, demonstrate the performance of MCDDPM alongside existing methods across various datasets. MCDDPM with $p=1$ outperforms existing methods on BraTS20, BraTS21 in terms of both Dice and AUPRC metrics. MCDDPM with $p=2$ exhibits even higher performance than $p=1$ achieving impressive Dice and AUPRC metrics on BraTS20 and BraTS21 datasets. Moreover, the reconstruction errors for MCDDPM ($p=1$) and MCDDPM ($p=2$) on the IXI dataset further validate their efficacy in image reconstruction. Qualitative results on a sample image from BraTS21 dataset provided in Fig. \ref{fig:qualitativeresults} show improved anomaly localization of MCDDPM. These findings underscore the superior performance of MCDDPM over existing methods across diverse datasets, suggesting its potential for various medical imaging applications. We conducted a separate analysis on the MSLUB data, calculating the dice score between the original masks and randomly generated binary masks. The resulting Dice score was approximately 4\%. Considering that our model and other models in Table \ref{tab:result} achieved dice score $\approx 5\%$, these slight improvements may be attributed to impact of random noise and do not reflect actual model efficacy. Hence interpreting results obtained for MSLUB dataset should be done carefully.

\section{Ablation Study}
We first explored the effect of different \( \lambda \) values on the reconstruction performance of the bridge network. By experimenting with \( \lambda \) values in $\{0.5, 1, 2\}$, we aimed to identify a suitable $\lambda$ value that maximizes the bridge network's reconstruction capabilities. Secondly, to evaluate the significance of the bridge network within our architecture, we conducted experiments by excluding it and subsequently assessing the model's ability to reconstruct. This step allowed us to quantify the contribution of the bridge network to the overall performance. Lastly, we delved into the role of conditional features, specifically examining the importance of integrating the context vector \( C \) in the denoising U-Net architecture. By comparing the performance of the model with and without this integration, we sought to determine the impact of conditional features on the model's reconstruction efficacy. All ablation studies were performed with $p=2$.

\begin{table}[!h]
	\centering
	\caption{Impact of various components and hyperparameters}
	\resizebox{\columnwidth}{!}{%
		\begin{tabular}{lcllllll}
			\toprule
			&\multicolumn{7}{c}{\textbf{Datasets}} \\      
			\cmidrule{2-8}
			& \multicolumn{1}{c}{IXI} & \multicolumn{2}{c}{BraTS20} & \multicolumn{2}{c}{BraTS21} & \multicolumn{2}{c}{MSLUB} \\
			\cmidrule{2-8}
			\textbf{Ablations}  & Reconstruction & Dice & AUPRC & Dice & AUPRC & Dice  & AUPRC \\
			& Error & & & & & & \\
			\midrule
			$\lambda$=0.5 & 0.01132  & 37.18 & 41.72 & 50.56 & 57.83  & 5.49 & 6.67 \\
			$\lambda$=1 & 0.01218 & 36.74 & 40.96 & 50.29 & 56.25 & 5.63 & 6.36 \\
			$\lambda$=2  & 0.01258 & 36.62 & 40.70 & 50 & 56.05 & 4.24 & 6.61 \\
			\midrule
			Without Bridge Network & 0.01363 & 33.53 & 35.39 & 44.43 & 45.24 & 5.13 & 6.28 \\
			Without Conditional features & 0.01494 & 32.85 & 37.32 & 44.25 & 48.18 & 5.72  & 6.41 \\
			\bottomrule
		\end{tabular}
	}
	\label{tab:result2}
\end{table}
The detailed findings from Table~\ref{tab:result2} indicate that \( \lambda=0.5 \) is a suitable choice for effective reconstruction of the bridge network facilitating more effective feature capture and representation, thereby enhancing Dice and AUPRC scores for the BraTS20 and BraTS21 datasets. Conversely, increasing \( \lambda \) values or removing the bridge network altogether leads to a decline in performance metrics across most datasets, highlighting the detrimental impact of over-regularization or the absence of critical architectural components. Additionally, the removal of conditional features, designed to enhance contextual information integration, results in declined performance in BraTS20 and BraTS21 datasets. 

\section{Conclusion}

 In this paper, we have proposed MCDDPM, that employs a tailored U-Net architecture combined with a bridge network to improve image reconstruction in denoising diffusion probabilistic models (DDPM). Two key features of our approach stand out, namely the enhanced bridge network capable of extracting multichannel information from different forms of noisy images, and the integration of a context vector into the denoising process of DDPM enabled by substituting the bottleneck self-attention layer with a cross-attention layer in the U-Net, with no extra trainable parameters. These modifications allow the model to effectively incorporate both enhanced input data representations and contextual information, improving its adaptability and performance. Our comprehensive evaluation against state-of-the-art competing methods on diverse datasets demonstrates MCDDPM's superiority in terms of reconstruction capability and anomaly detection efficacy. 

\section*{Acknowledgments}
The authors thank Technocraft Centre for Applied Artificial Intelligence (TCA2I), IIT Bombay, for their generous funding towards this project.

\renewcommand{\bibfont}{\scriptsize}
\bibliographystyle{IEEEtranN}  
\bibliography{refrences}

\end{document}